\documentclass[journal]{IEEEtran}
\usepackage{cite}
\usepackage{algorithmic}
\usepackage{array}
\usepackage{amsmath,amsfonts,amssymb}
\usepackage{graphicx}
\usepackage{paralist}
\usepackage[colorlinks=true, allcolors=blue]{hyperref}
\ifCLASSOPTIONcompsoc
  \usepackage[caption=false,font=normalsize,labelfont=sf,textfont=sf]{subfig}
\else
 \usepackage[caption=false,font=footnotesize]{subfig}
\fi
\usepackage{stfloats}
\usepackage{url}

\begin{document}
\title{Performance and control strategy of an integrated tunable laser with a single intra-cavity AMZI filter }

\author{Martin Skënderas,~\IEEEmembership{Student Member,~IEEE,}
        Pablo Marin-Palomo,
        Spencer W. Jolly,~\IEEEmembership{Member,~IEEE}
        and~Martin Virte,~\IEEEmembership{Member,~IEEE}
\thanks{
This work was supported by Fonds Wetenschappelijk Onderzoek (FIONA, project number G029619N, COLOR’UP, project number G0G0319N, REFLEX, project number 1530318N).

Martin Skënderas, Pablo Marin-Palomo and Martin Virte are with the Brussels Photonics (B-PHOT), Vrije Universiteit Brussel, Department of Applied Physics and Photonics, Pleinlaan 2, B-1050 Brussels, Belgium.(e-mail:martin.skenderas@vub.be; martin.virte@vub.be)

Spencer W. Jolly is with the OPERA-Photonique, Université Libre de Bruxelles,
50 Av. F. D. Roosevelt, CP 194/5, B-1050 Bruxelles, Belgium.}
}

\markboth{}
{Shell \MakeLowercase{\textit{et al.}}: Bare Demo of IEEEtran.cls for IEEE Journals}

\maketitle
\begin{abstract}
Asymmetric Mach-Zehnder interferometers (AMZIs) can, in principle, enable continuous wavelength tuning of a laser when used as an intra-cavity filter. Their simplicity and good compatibility with generic foundry platforms are major advantages. However, the difficulty to develop a well-defined and robust control strategy is an important drawback which restricts the use-cases of these tunable lasers.
Here, we make an in-depth investigation of the tunability properties of a laser including a single-stage AMZI in its cavity. We find that due to imperfections of Electro-Optic Phase Modulators (EOPMs), the dependence of the phase variation with the applied voltage is not linear. Because integrated EOPMs cannot be individually calibrated, these nonlinearities prevent a precise and independent tuning of the phase and amplitude of the AMZI transfer function, and thus continuous tuning cannot be reliably achieved. To overcome this issue, we propose a refined control strategy which allows for semi-continuous tuning. With this approach, we demonstrate a piece-wise continuous tuning of the emission wavelength by taking advantage of the coupling between amplitude and phase in the AMZI response. With our refined control strategy, we achieve tuning of the emission wavelength over the full free spectral range (FSR) of the AMZI. 
\end{abstract}


\IEEEpeerreviewmaketitle
\section{Introduction}
\IEEEPARstart{I}{ntegrated} wavelength-tunable laser sources have become increasingly important due to their implementation in a wide variety of applications such as communication systems, spectroscopy, sensing and metrology \cite{duarte, 1266694, Buus:06, 8047312}. Precise tuning of the wavelength is required in these applications e.g., to accurately record the shape of the resonance lines in spectroscopy or precisely locate the position of the Bragg wavelength in fiber Bragg gratings (FBGs). Depending on the intended application the required precision of the wavelength tuning varies from a few \,MHz to a few \,GHz. 

Over the last years, several types of wavelength-tunable lasers have been developed based on different wavelength filtering mechanisms and designs such as ring resonators \cite{998697, app7070732, 4811964, 7111228}, distributed Bragg reflectors (DBR) \cite{1300621, Andreou:19, Happach:17} and coupled cavity lasers \cite{DAgostino:15, 6522125, doi:10.1063/1.3514247}. Among the various concepts, wavelength-tunable lasers based on asymmetric Mach-Zehnder interferometer (AMZI) as the wavelength selective intra-cavity filter are attractive since, in principle, the AMZI enables a continuous wavelength tuning over a wide spectral range and its performance is robust to the tolerances of the fabrication processes. However, if the AMZI is the only wavelength-selective mechanism of the laser, more than one AMZI is needed to obtain single mode operation. Tunable AMZIs as wavelength filters have been employed by Latkowski et al. \cite{7305755} inside the cavity of a ring laser ensuring a record tuning range of $74.3$\,nm around $1525$\,nm as well as to obtain tunable sources operating at $2$\,$\mu$m \cite{Latkowski:16_optica}. In  these designs, three AMZIs in series \cite{7305755} or two nested AMZIs \cite{Latkowski:16_optica} have been used.   

Despite the wide and thus attractive tuning range, widely reported in literature, no accurate control approach has been derived yet. A precise and efficient control represents a technique through which the dependence of the filter's transfer function on the filter parameters is fully determined and accurately connected with the emission wavelength. 
 This issue is first tackled in Ref. \cite{10.1117/12.2509572} where attempts to exploit the serially cascaded AMZI configuration \cite{7305755} for applications in optical coherence tomography (OCT) have been made. This brought the need to calibrate the response of the emission wavelength on the phase modulators voltage amplitude, in order to tune the laser in a step-wise manner with $1$\,GHz accuracy. A three step calibration method with progressive wavelength accuracy has been proposed \cite{10.1117/12.2509572}, although the tuning of the coarse and medium AMZI alone leads to wavelength gaps within the tuning range. Moreover, the electrical cross-talk between the phase modulators limits the tuning accuracy to $8$\,GHz. Recently, an optimized control strategy for the wavelength calibration of an improved design configuration has been reported \cite{10.1117/12.2610120}. This strategy re-calibrates the fine filter and the cavity mode position every 24h for every coarse and medium filter control setting configuration. A table with uniform wavelengths with $10$\,GHz  frequency steps has been obtained. Unfortunately, a sequential selection of the longitudinal modes cannot be achieved and wavelength gaps as wide as $2.5$\,nm are still present. Hence, a technique giving access to the entire spectral range of an AMZI-based laser with continuous wavelength tuning remains to be developed. 
 
 In this paper we investigate the control mechanism of a single stage AMZI-tuned laser in order to gain further insight on its tuning limitations and control challenges. Here, the AMZI is combined with a DBR, which eliminates the need of additional intra-cavity AMZIs, while being easily controlled and more robust. This approach however limits the tuning range. We show that the simple tuning method of reverse biasing the phase modulators, results in wavelength gaps within the free spectral range (FSR) of the AMZI. Moreover, the comprising phase modulators have different dependencies with the applied voltage, hence, a correction is needed. Thus, we propose a control strategy that takes advantage of the phase difference between the AMZI arms and demonstrate that a semi-continuous tuning with wavelength tuning resolution up to $0.1$\,pm can be achieved, covering the entire FSR of the AMZI and $89.5$\% of the laser's tuning range. In addition, we examine the spectral properties, frequency noise power density, and the wavelength switching time of the laser. 
 
 The paper is organized as follows. In section \ref{sec:design}, we present details on the cavity design, design consideration and working principle of the AMZI. In section \ref{sec:characterization}, the experimental results on the current-light-voltage characteristics and the spectral quality are presented. The power spectral density of the frequency noise and the wavelength switching speed are determined and compared with the values reported in literature.  In section \ref{sec:tuning challenges}, challenges for a simple step-wise and sequential tuning of the emission wavelength are highlighted and in section \ref{sec:tuning strategy}, a control strategy that enables semi-continuous tuning is presented. 
  
\section{Design considerations of an intra-cavity AMZI based tunable laser}
\label{sec:design}
\begin{figure}
  \begin{center}
  \begin{tabular}{@{}c@{}}
    \includegraphics[height=4cm]{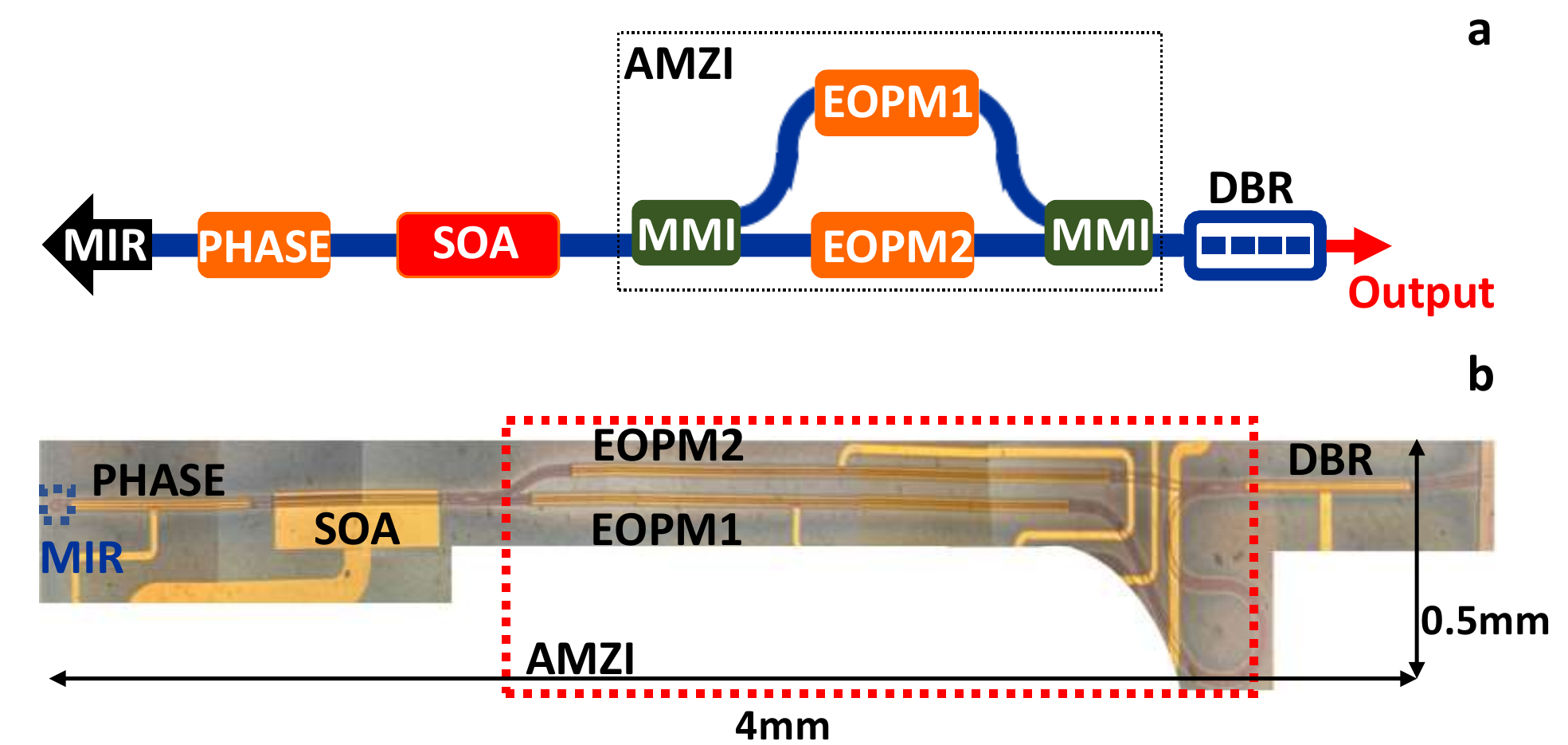}
  \end{tabular}
  \end{center}
     \begin{center}
  \begin{tabular}{@{}c@{}}
     \includegraphics[width=\linewidth]{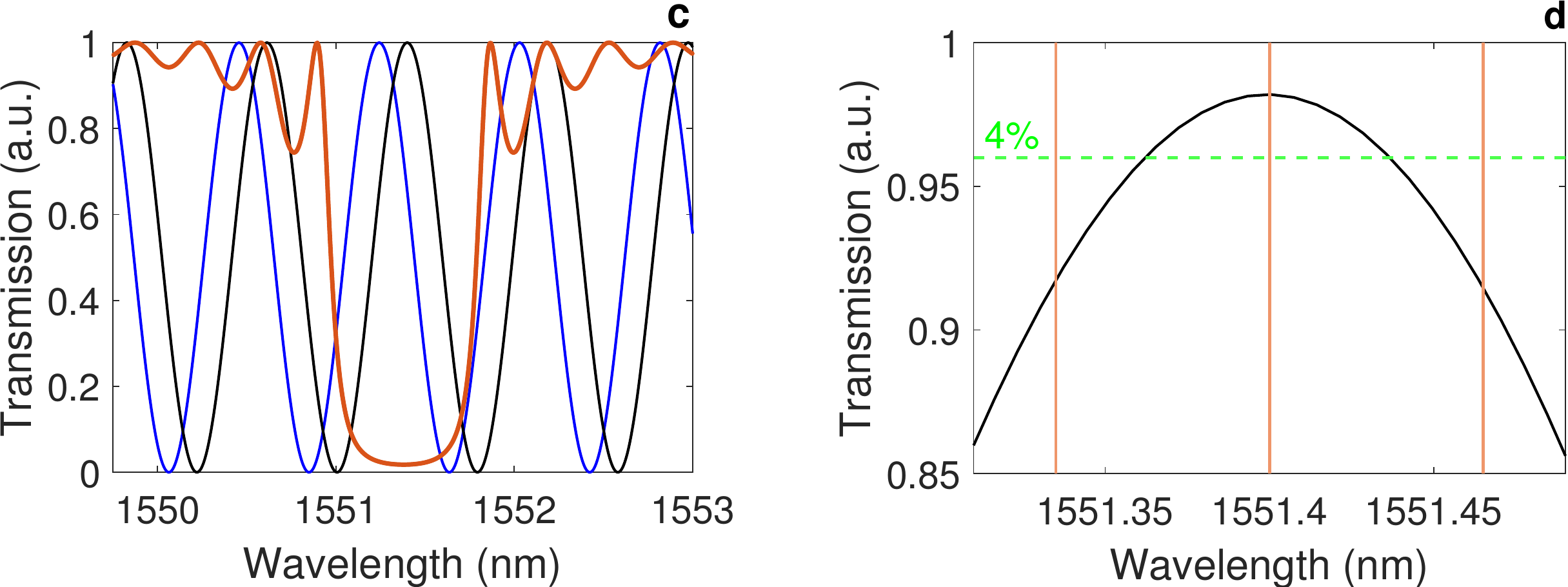} \\[\abovecaptionskip]
  \end{tabular}
  \end{center}
  \caption{ (a) Schematic of the tunable integrated laser with an intra-cavity AMZI. The MIR and the DBR mirror form the cavity with an SOA and a phase section. The coupling to the AMZI is implemented using $2\times1$ MMIs. (b) Microscope image of the fabricated laser. The chip is mounted on an aluminium platform and wire-bonded to circuit boards. (c) Calculated transmission profile of the AMZI filter with total FSR of $0.66$\,nm, with blue for $V_\textrm{EOPM1}$ and $V_\textrm{EOPM2}$ equal to $0$\,V and with black for $V_\textrm{EOPM1}=0$\,V and $V_\textrm{EOPM2}$ equal to -$2$\,V, superimposed to the transmission profile of a $500$\,$\mu$m long DBR in orange. (d) Close-up of the combined transmission profile of the AMZI and the reflectivity of the DBR (black) and longitudinal cavity modes (orange) resulting in single mode selection.}
  \label{fig:Design}
\end{figure}

The AMZI is built with two $1.8$\,mm long electro-optic phase modulators (EOPMs), one in each arm, and two $1\times2$ multimode interference reflectors (MMIs), as splitter and coupler. The asymmetry is created by introducing an optical path length imbalance $\Delta L$ equal to $965$\,$\mu$m. The theoretical transmission of an individual AMZI, blue curve in Fig. \ref{fig:Design}(c), has a periodic behavior in the frequency domain, with a cosine-squared transmission profile. The free spectral range (FSR), defined as the spacing between two successive transmitted optical intensity maxima or minima, of the AMZI is $83.1$\,GHz ($0.66$\,nm) estimated by the relation: $FSR=c/n\Delta L$ where c is the speed of light in vacuum and $n$ the group refractive index of the deeply etched waveguide. The asymmetry introduces a different optical path-length in each of the branches, hence a relative phase difference given by\cite{Chtcherbakov:98,el_shamy_modelling_2022}: 
\begin{align}
\Delta\Phi=\frac{2\pi n \Delta L}{\lambda}
\label{eq:phase_difference}
\end{align}
where $\lambda$ is the wavelength in vacuum. A change of the optical phase by $2\pi$ between the arms tunes the filter over one full FSR of the AMZI. The voltage necessary to induce this phase change between the AMZI arms is denoted as $V_{2\pi}$, and is experimentally estimated to be $\approx 10.4$\,V. The resulting electric field after the AMZI is mathematically written as:
\begin{multline}
    E_{out}=E_{in}\cos{\left[2\pi\left(\frac{V_\textrm{EOPM1}-V_\textrm{EOPM2}}{2V_{2\pi}}-\frac{n\Delta L}{2\lambda}\right)\right]}\\ 
    \times e^{-j2\pi\frac{V_\textrm{EOPM1}+V_\textrm{EOPM2}}{2V_{2\pi}}} 
    \label{eq:intensity}
\end{multline}
where $E_{in}$ is the input field amplitude and $V_\textrm{EOPM1}$ and $V_\textrm{EOPM2}$ the voltage applied in EOPM1 and EOPM2 respectively. Hence, the AMZI transmission maxima can be tuned by applying a reverse bias voltage in one of the EOPMs, or both EOPMs simultaneously, as shown with black in Fig. \ref{fig:Design}(c).

The transmission of an individual AMZI in combination with the fundamental mode structure of the laser cavity is not sufficient to achieve single mode operation as there will be selected modes for every AMZI transmission maxima. For that, we use a DBR to select only one of the transmission maxima as illustrated with orange in Fig. \ref{fig:Design}(c). To choose one of the transmission maxima, in the design procedure, we made a trade-off between the $\Delta L$ of the AMZI and the width of the DBR stop-band in order to allow for only one maximum of the transmission spectrum of the AMZI within the DBR stop-band. A larger imbalance $\Delta L$ between the AMZI arms would mean that other AMZI transmission peaks might appear in the DBR stop-band and, a smaller imbalance $\Delta L$ would compromise the side-mode suppression ratio (SMSR) as the filtering function will be less selective. The stop-band width of the DBR is approximately $1.2$\,nm which means that the cavity modes that fall at the edge and outside of this band will be greatly suppressed. On the other hand, the length choice of the DBR is a compromise between its reflectivity and stop-band since a higher stop-band theoretically will result in a higher tuning range but also lower output power of the laser. We use a $500$\,$\mu$m long DBR, and its reflectivity is predicted to be around $90$\%. The wavelength selection is thus realised by combining one AMZI with a DBR. In order to obtain a good side-mode suppression ratio (SMSR), the proposed filter should ensure a sufficient mode selectivity, at least $4$\% transmission difference between the selected wavelength and the other competing modes \cite{7305755}. A close-up of the combined cavity mode selection in black, shown in Fig. \ref{fig:Design}(d), overlapped with the longitudinal cavity modes around the transmission peak demonstrate the analysis of the losses corresponding to the neighboring modes of the selected peak. With our set of parameters only one cavity mode will be selected. 

To form the laser cavity one multimode interference reflector (MIR) is used in the rear end of the laser. The laser also includes a $500$\,$\mu$m semiconductor optical amplifier (SOA) which provides the optical gain, and a $500$\,$\mu$m long EOPM as a phase section. Reverse biasing the phase section enables in-line cavity phase adjustments and therefore the translation of the cavity modes which can aid to continuously tune the laser. The overall cavity length is approximately $4860$\,$\mu$m which results in a longitudinal mode spacing of $\approx 65$\,pm. The proposed design is shown schematically in Fig. \ref{fig:Design}(a). The laser was fabricated using a commercially available active-passive InP-based integration technology in the framework of a multi-project wafer run by Smart Photonics. A microscopic image of the laser is shown in Fig. \ref{fig:Design}(b). Its footprint is less than $4\times0.5$\,mm$^2$.

\section{Laser characterization}  
\label{sec:characterization}
\subsection{Spectral performance and tuning capability}
The fabricated chip is mounted on an aluminium platform and wire-bonded to circuit boards for ease of control and characterization. The waveguide from the laser output is angled with respect to the chip facet to suppress back-reflections to the laser cavity.  The laser output light was coupled out of the chip using a single-mode lensed fibre. The laser was characterized at $18$\,$^{\circ}$C, where the emitted power is the highest, using a thermistor controlled via a temperature controller (Thorlabs PRO800/ITC8052). The SOA and DBR current is provided by a laser diode controller (Thorlabs PRO800/ITC8052) and the output optical power is monitored by a power meter. The two EOPMs of the AMZI are reversed biased from $0$\,V to -$10$\,V and, the phase section from $0$\,V to -$5$\,V using a voltage source module (NI9253). $1000$ quasi-random voltage combinations are generated and applied to the EOPMs. The optical spectra are recorded for each case with a $5$\,MHz resolution bandwidth using an optical spectrum analyzer (APEX 2083A).

\begin{figure} 
   \begin{center}
   \begin{tabular}{c}
   \includegraphics[width=1.03\linewidth]{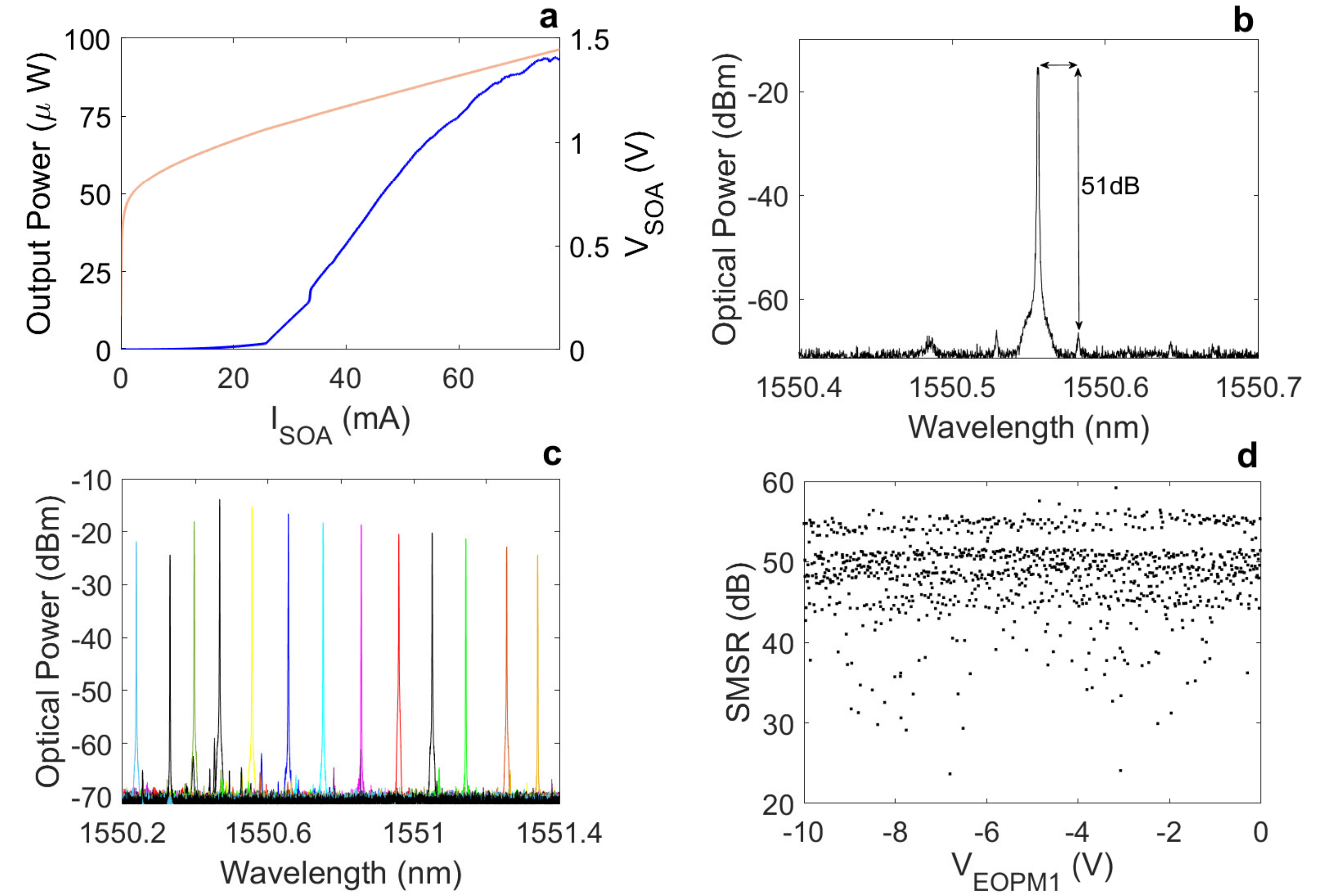}
   \end{tabular}
   \end{center}
   \caption[SpectralCharacterozation] 
   { \label{fig:SpectralCharacterization} 
(a) Optical output power in fibre (blue) and voltage (orange) of the laser as a function of SOA injection current at $18$\,$^{\circ}$C. (b) Emission spectrum of the integrated laser recorded with the APEX optical spectrum analyzer with $51$\,dB SMSR. (c) 13 overlapped optical spectra illustrating a tuning range of $1.2$\,nm with $45$\,mA SOA current. (d) The SMSR of the laser for different lasing wavelength as a function of $V_\textrm{EOPM1}$, resulting from the different reverse biased voltage combinations of $V_\textrm{EOPM1}$, $V_\textrm{EOPM2}$ and $V_\textrm{Phase}$.}
   \end{figure}

In Fig. \ref{fig:SpectralCharacterization}(a) the optical output power and the voltage as a function of the injection current are shown. The threshold current is about $25$\,mA and the output optical power reaches about $100$\,$\mu$W at $80$\,mA SOA current. This is the optical power in the lensed fibre to which we should add the coupling losses from the chip facet to the lensed fiber and which are estimated to be around $3$-$4$\,dB. The emitted power of the laser is relatively low which likely results from the high reflectivity of the DBR at the output of the laser. 

In Fig. \ref{fig:SpectralCharacterization}(b) the output optical spectrum is presented, with $5$\,MHz resolution bandwidth, for the voltage combination $V_\textrm{EOPM1}=\,$-$8.066$\,V, $V_\textrm{EOPM2}=\,$-$7.586$\,V and $V_\textrm{Phase}=\,$-$1.976$\,V. The closest neighboring modes to the lasing mode are barely visible above the noise floor of the instrument. This is in line with the design considerations discussed above. The laser is single mode and the SMSR is about $51$\,dB. $13$ optical spectra across the tuning range of the laser operated at $45$\,mA SOA current are shown in Fig. \ref{fig:SpectralCharacterization}(c). The tuning range reaches up to $1.2$\,nm, from $1550.2$\,nm to $1551.4$\,nm defined by the DBR band-stop. From all the recorded spectra the SMSR values are extracted and are plotted in Fig. \ref{fig:SpectralCharacterization}(d) as function of $V_\textrm{EOPM1}$. Each point in this plot correspond to different voltage combinations of $V_\textrm{EOPM1}$, $V_\textrm{EOPM2}$ and $V_\textrm{Phase}$. The SMSR of the optical spectra exceeds $20$\,dB across the whole tuning range where the most common value is between $45$\,dB and $60$\,dB.

\subsection{Intrinsic linewidth}
Intrinsically, the broadening of the laser linewidth is caused by the coupling of the spontaneous emission into the oscillating mode, leading to spectrally white frequency noise and to a Lorentzian laser line shape with a full-width at half maximum (FWHM) $\Delta f_L$. Typically, the frequency noise spectrum of a laser is also composed of the flicker noise and random-walk frequency noise \cite{6466346}. 
The frequency-modulation (FM) noise spectrum contains the complete statistical characteristics of the frequency noise of the laser. The FM spectrum is obtained from the power-spectral-density (PSD) function of the instantaneous optical frequency fluctuations. If the FM-noise spectrum is recorded for a short enough time \cite{Marin-Palomo:19}, the impact of flicker and random-walk frequency noise can be excluded and the intrinsic Lorentzian linewidth is directly proportional to the constant value of the FM noise spectrum \cite{Kikuchi:12}. 

\begin{figure}
  \centering
  \begin{tabular}{@{}c@{}}
    \includegraphics[height=2.5cm]{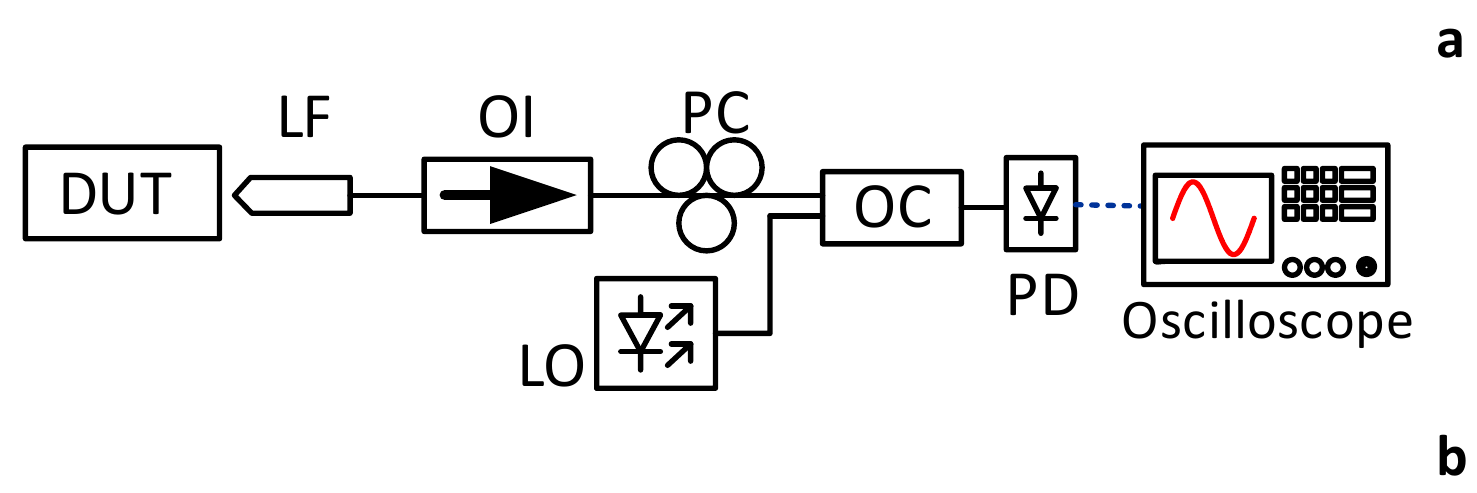} 
  \end{tabular}
  \vspace{\floatsep}
  \begin{tabular}{@{}c@{}}
    \includegraphics[width=0.7\linewidth,height=3cm]{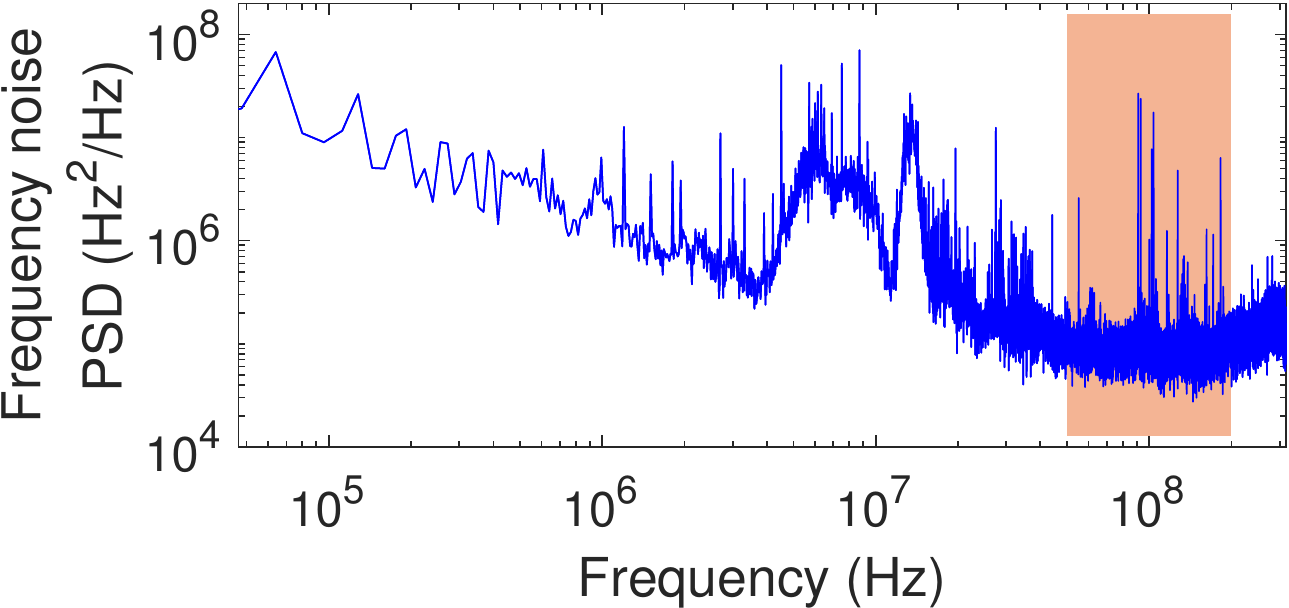} \\[\abovecaptionskip]
  \end{tabular}
  \caption{ Linewidth measurement and phase-noise characterization for the tunable laser. (a) Experimental setup: DUT: device under test. LF: lensed fiber. OI: optical isolator. PC: polarization controller. LO: local oscillator. OC: optical coupler. PD: photodiode. (b) Power spectral density of the frequency noise of the laser. The orange region indicates the flat part of the frequency noise which when multiplied by $\pi$ yields the intrinsic Lorentzian linewidth.}
  \label{fig:Linewidth}
\end{figure}

To obtain the FM-noise spectrum and characterize the laser phase noise, we use the setup shown in Fig. \ref{fig:Linewidth}(a). The emitted light is sent through a polarization controller (PC) and then superimposed with a local oscillator (LO) laser in an optical coupler. The LO is a high-quality external-cavity laser (ECL) (Keysight N7776C) with an intrinsic optical linewidth smaller than $10$\,kHz. At the output, a photodetector (Thorlabs RXM42AF) generates a beat signal at an intermediate frequency around $2$\,GHz defined by the detuning of the LO and the wavelength of interest. Then the beat signal is recorded using a high definition oscilloscope (Teledyne Lecroy WavePro 804HD) which works as an analog-to-digital converter (ADC) with a sampling rate of $20$\,GS/s. For this experiment we have recorded $150$\,MSa. The frequency noise spectrum is then calculated by taking the Fourier transform of the extracted instantaneous frequency noise fluctuations and is plotted in Fig. \ref{fig:Linewidth}(b).

The spectral density of the frequency noise becomes white approximately between $5$\,MHz and $20$\,MHz  where other frequency noise effects mentioned above are neglected and the frequency noise PSD is flat. To find the frequency noise power spectral density level we average the noise level in this region which is highlighted in Fig. \ref{fig:Linewidth}(b) and obtain $121$\,Hz$^2$/Hz. When the level of the flat part of the singe-sided frequency noise is multiplied by $\pi$ we obtain the Lorentzian intrinsic linewidth of $381$\,kHz.
The value found for the Lorentzian linewidth is in the same order of magnitude as for previously reported lasers in this integration platform \cite{Andreou:19, 8894462}.

\subsection{Wavelength switching time}
\begin{figure}
  \centering
  \begin{tabular}{@{}c@{}}
    \includegraphics[height=3.5cm]{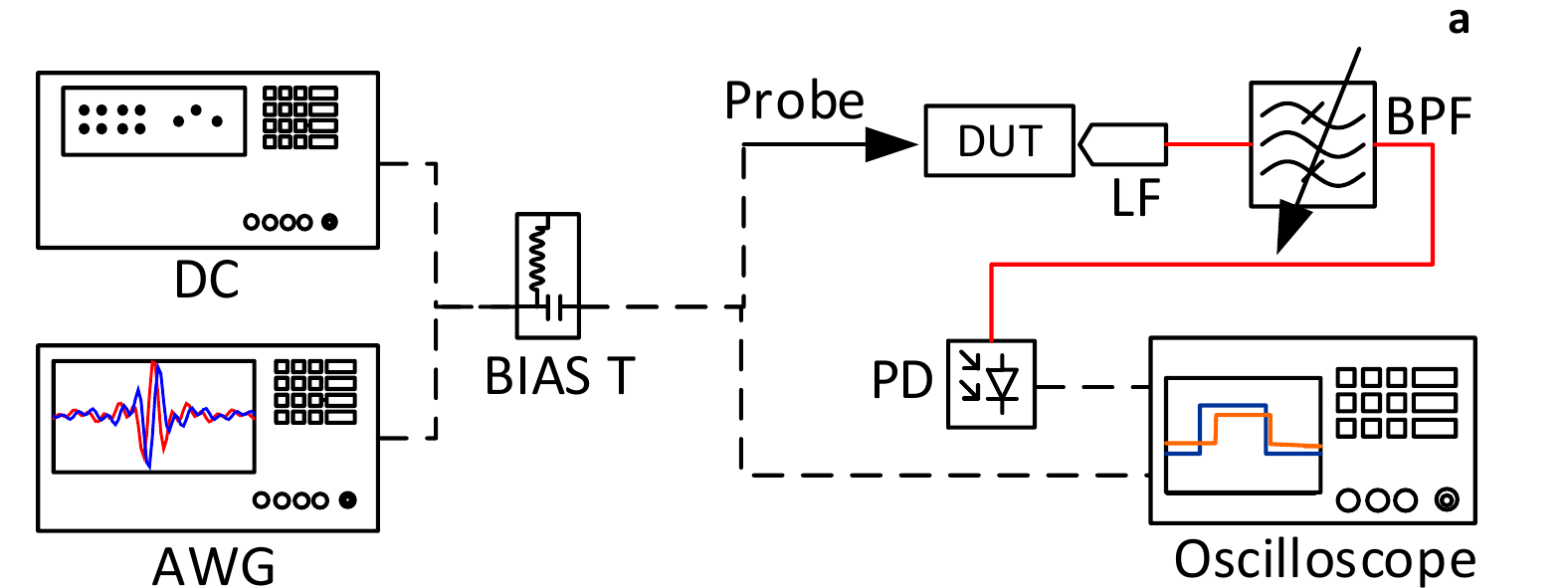} 
  \end{tabular}
  \vspace{\floatsep}
  \begin{tabular}{@{}c@{}}
    \includegraphics[width=\linewidth]{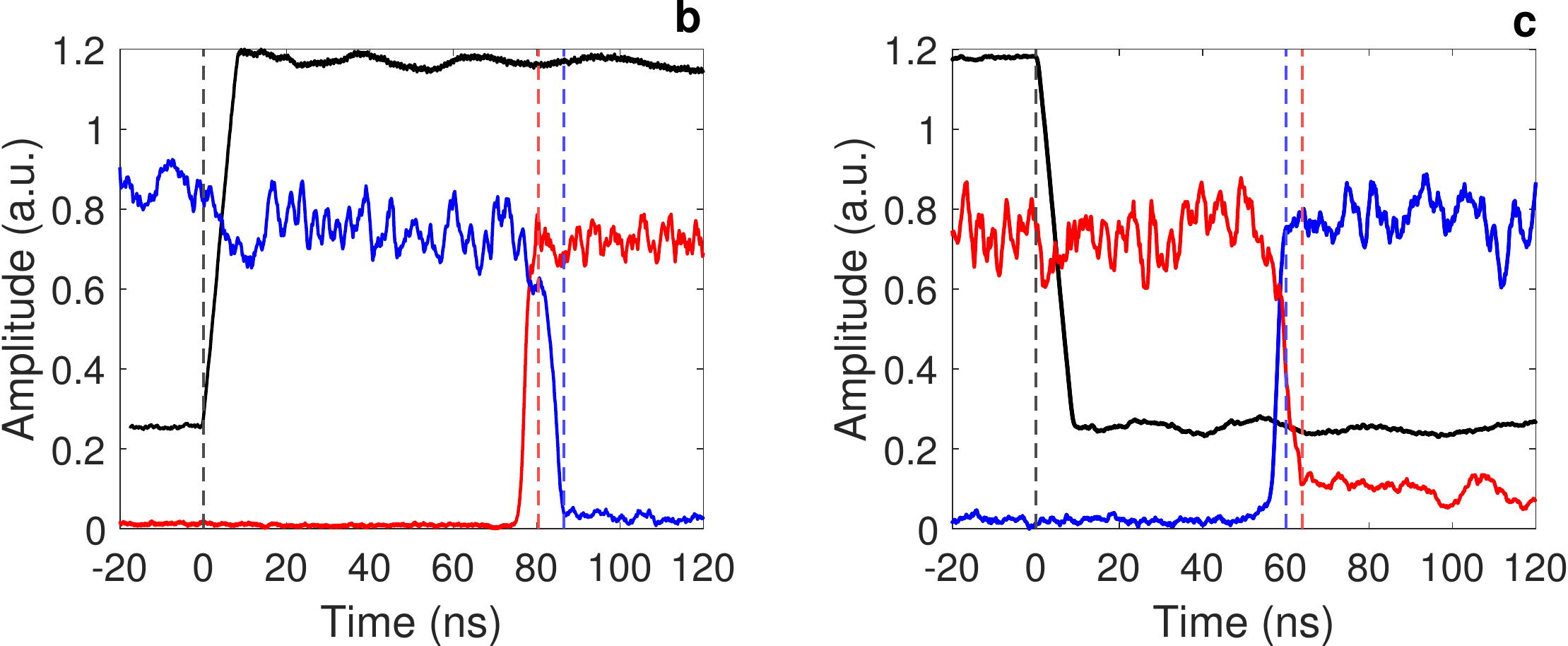} \\[\abovecaptionskip]
  \end{tabular}
  \caption{Wavelength switching speed for two wavelength positions within the tuning range. (a) Experimental setup: AWG: arbitrary waveform generator. DC: voltage source. DUT: device under test. PD: photodiode. LF: lensed fiber. BPF: tunable bandpass filter. Switching speed measurements. (b) Amplitude of the signal: In black of the rising trigger edge, in red $\lambda_1$ being turned on, and in blue $\lambda_2$ being turned off. (c) Amplitude of the signal:  In black of the falling trigger edge, in red $\lambda_1$ being turned off, and in blue $\lambda_2$ being turned on. The trigger signal serves as a reference and is shifted vertically for clarity. The black dotted line indicate the reference signal, and colored dotted lines the $10$\% and $90$\% values of the wavelengths being turned off and on, correspondingly.}
  \label{fig:wav_switching}
\end{figure}

The switching time between the two wavelengths is determined as the time required for the power associated to the original wavelength to decrease below $10$\% of the initial power i.e., switch-off, and until the power associated to the targeted wavelength reaches $90$\% of the final power level i.e., switch-on. Here, we chose two neighboring longitudinal modes within the tuning range: $\lambda_1$\,=\,$1550.25$\,nm and $\lambda_2$\,=\,$1550.31$\,nm.
To study the wavelength switching speed, we used the experimental setup shown in Fig. \ref{fig:wav_switching}(a). It consists of an arbitrary waveform generator (Keysight M9502A) that generates a step function which is added through a bias T to a DC signal, generated by a voltage source (Agilent E3646A). The EOPMs integrated in the AMZI are not optimized for radio-frequency (RF) operation, we therefore use a standard DC probe to contact them. The emitted light is coupled into the lensed fibre after which an optical bandpass filter (EXFO XTM-50) selects one of the two wavelengths for measurements using an oscilloscope (Tektronix CSA 7404) with a built-in photo-diode. All the fibers, coaxial cables, and devices introduce time delays. This means that the time it takes the signal to propagate towards the oscilloscope is longer compared to the reference signal, therefore, the time delay introduced by the setup has to be taken into account. The difference in signal time delay between the trigger and the signal was found to be $24.5$\,ns and is corrected in the experimental results. This measurement technique allows to monitor only one wavelength at a time, so switching is repeated after tuning the bandpass filter to the other wavelength.  

We applied a step function with a pulse width of $250$\,ns, rise and fall time of $10$\,ns and a peak-to-peak amplitude of $0.5$\,V. A DC voltage of $6.5$\,V is applied to the EOPM2 of the AMZI via the bias T. The experimental results are shown in Fig. \ref{fig:wav_switching} for the rising and falling trigger edge in (b) and (c), respectively. The trigger signal is shown in black and is vertically shifted for clarity, and the amplitude of the signal associated to the wavelengths $\lambda_1$ and $\lambda_2$ are shown in red and blue, correspondingly. The vertical black dotted line indicates the trigger signal as the reference, and the coloured dotted lines indicate the thresholds of $10$\% and $90$\% for the wavelengths being turned on and off. 
The amplitudes have been normalized to allow an easy comparison. We find response times for $\lambda_1$ of around $80$\,ns to switch-on and about $60$\,ns to switch off, and for $\lambda_2$ around $64$\,ns to switch-on and $86.6$\,ns to switch-off. We notice that for the falling trigger edge the switching times are smaller compared to the rising trigger edge. The transition time itself takes a few \,ns. With a similar design but including EOPMs optimized for RF operation --- which can reach a bandwidth of up to $25$\,GHz on the generic platform we use --- we expect the switching to be limited only by the laser dynamics itself and thus to potentially be significantly faster. High modulation speed has been reported for this integration platform \cite{Smit_2014}, which can enable fast wavelength switching well into the \,ns regime \cite{4456801, 4838803}. It has been shown that the intra-cavity cascaded AMZI design the wavelength switching time of two non-sequential wavelength is around $100$\,ns \cite{10.1117/12.2555620}, which is slightly higher than what we measure. Considering that, during the wavelength switching the power of the original mode decreases while the power of the targeted mode increases, for a continuous, sequential tuning, the best strategy is to switch the filter to neighboring longitudinal cavity modes. We have selected two sequential modes, with the argument that there is already light at the targeted wavelength and it can be amplified faster. If the filter switches to a random mode within the tuning range of the laser, the light field build up in the target mode is expected to take longer and hence, longer switching times. 

\section{Challenges of a robust control for a AMZI-based laser} 
\label{sec:tuning challenges}
As explained in section \ref{sec:design}, our laser is designed such that it allows for the selection of one mode at a time enabling single mode emission. Here, our intention is to define the required voltages to be applied to EOPM1 and EOPM2 in order to select consecutive longitudinal modes within the DBR stop-band with a relatively simple strategy. During this experiment the sum of the voltages $V_\textrm{EOPM1}$ and $V_\textrm{EOPM2}$ is kept constant in each measurement step, thus, the phase term in Eq.\,\ref{eq:intensity} should ideally be constant, which means that the position of the longitudinal cavity modes should remain unchanged. This would ensure that the amplified mode i.e., the emission wavelength of the laser depends solely on the filter and the selected cavity mode only on the value $\Delta V=V_\textrm{EOPM1}-V_\textrm{EOPM2}$. Thus, we vary the voltage in each EOPM in opposite directions i.e., $V_\textrm{EOPM1}$ from -$10$\,V to -$5$\,V and $V_\textrm{EOPM2}$ from -$5$\,V to -$10$\,V, in steps of $0.005$\,V using a voltage source module (NI9253). The voltage of the phase section is set at $0$\,V, and is kept unchanged throughout the experiment. The optical spectra are recorded using an optical spectrum analyzer (OSA) (APEX 2083A) with a resolution of $1.12$\,pm. 

   \begin{figure} 
   \begin{center}
   \includegraphics[width=0.88\linewidth]{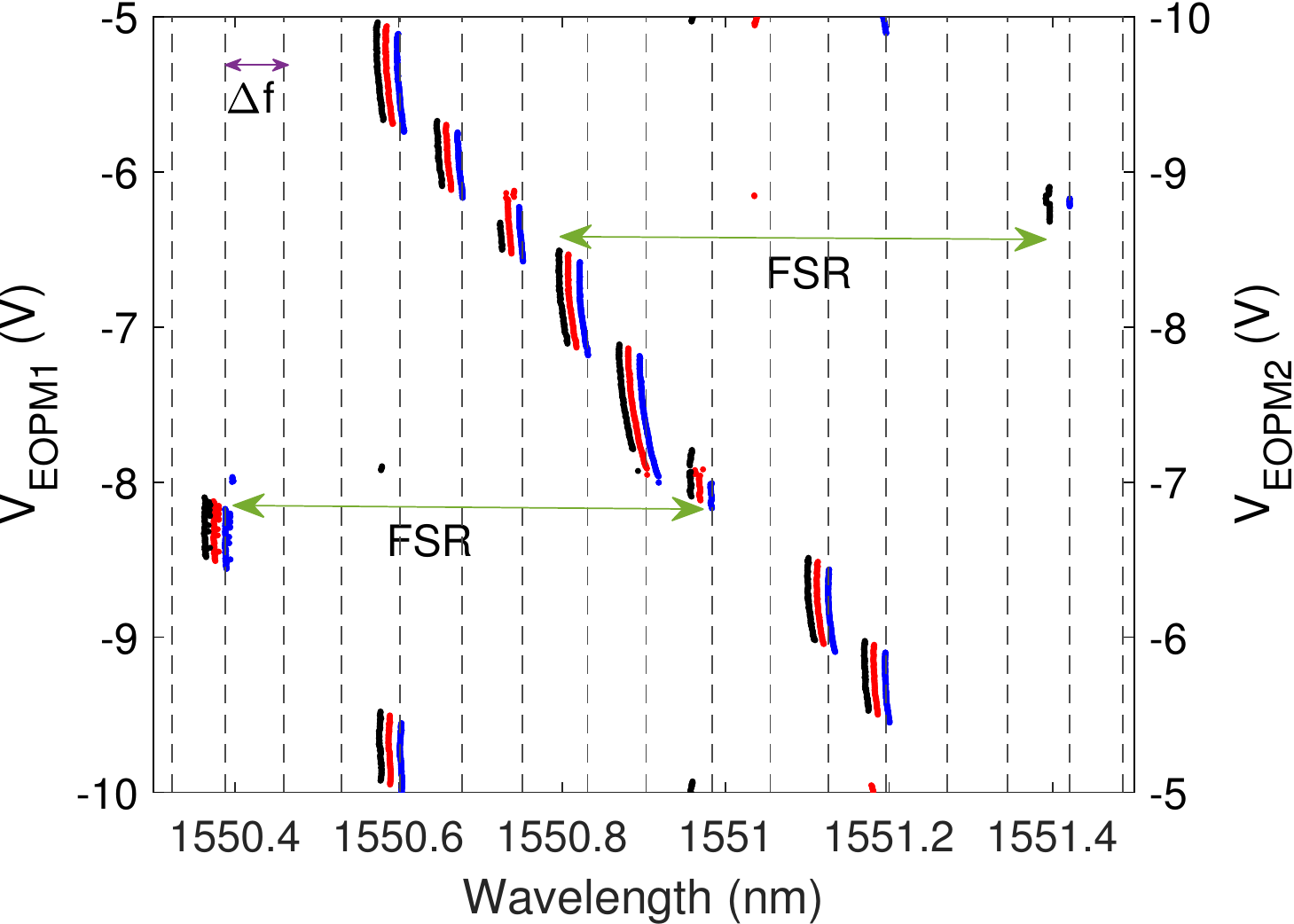}  
   \end{center}
   \caption[ReverseTuning] 
   { \label{fig:Revrese_tuning} 
The emission wavelength of the laser when the voltages of the AMZI EOPMs are switched in opposite directions and their sum is kept constant at every step, for different values of the phase section voltage; $0$\,V in blue, -$4$\,V in red and -$8$\,V in black. The black dotted vertical lines show the position of the longitudinal cavity modes separated by $\Delta f$, and the green arrows indicate the positions where the wavelength changes by the FSR of the AMZI. }
   \end{figure} 
The emission wavelength is extracted from the OSA measurements and the results are plotted in blue in Fig.\,\ref{fig:Revrese_tuning}. Several longitudinal modes separated by $\Delta f$ are selected, we identify them with black doted lines. This simple tuning approach enables a step-like wavelength tuning where the EOPM voltages are switched continuously with the conditions mentioned above. We notice some features resulting from this control approach: 
  \begin{inparaenum}[i)]
    \item when the voltages are varied continuously and the resulting transfer function of the AMZI is in the edge of the DBR transmission, the following AMZI transmission will be on the other edge of the DBR which results in sudden wavelength jumps equal to the FSR of the AMZI, as shown with the green arrows in Fig.\,\ref{fig:Revrese_tuning}. This behavior constrains the sequential selection of all the longitudinal modes within the FSR of the AMZI. A simple way to prevent it, is to exclude the voltage combinations resulting in such sudden jumps, or alternatively fine tune the design parameters to increase the AMZI transmission period or decrease the DBR stop-band; 
   \item there are gaps between discrete points within the wavelength tuning range which means that not all possible cavity modes can be amplified, and the wavelength tuning resolution is also defined by the cavity mode spacing, $\Delta f$. The existence of the gaps within the tuning range is an inherent feature linked to the AMZI filter performance and this control approach. However, if more cavity modes would be present, they would potentially be selected by the AMZI filter, in turn affecting the wavelength tuning resolution. For that we shift the cavity modes by varying the voltage of the phase section ($V_\textrm{Phase}$) and then tune the AMZI in the same way as in the previous experiment. The results are plotted in Fig.\,\ref{fig:Revrese_tuning} with red for $V_\textrm{Phase}=\,$-$4$\,V and with black for $V_\textrm{Phase}=\,$-$8$\,V. The emission wavelength changes in roughly the same trend as for the case of $V_\textrm{Phase}=0$\,V. With higher voltages or a longer phase section, the wavelength range between cavity modes would be attainable enabling a continuous wavelength tuning. But, that would come at the expense of the simplicity of the control mechanism since the three distinct voltages would need to be adjusted at every step;
   \item for the voltage combinations that select one cavity mode it appears that the resulting wavelength peaks follow a slope which means that the modes are slightly shifted indicating the effect of an extra phase between the AMZI arms.
  \end{inparaenum}

\begin{figure} 
   \begin{center}
   \includegraphics[width=\linewidth]{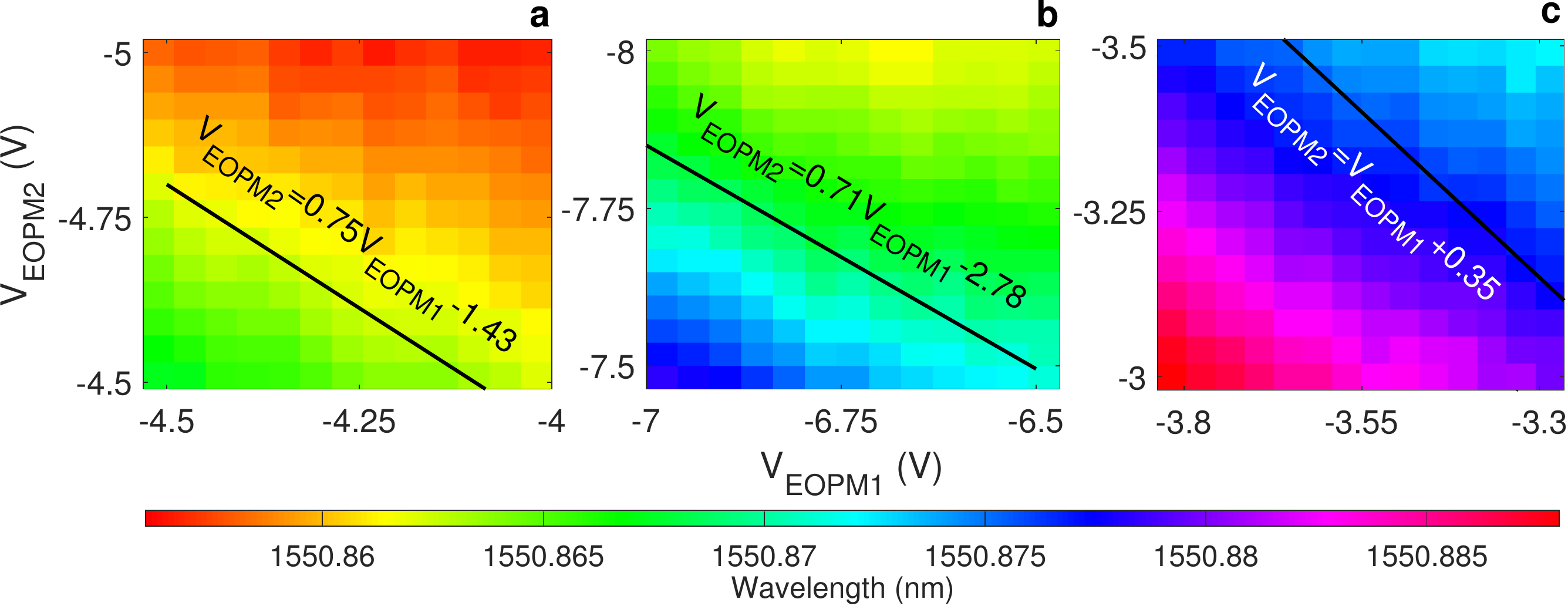}  
   \end{center}
   \caption[Gradient] 
   { \label{fig:Gradient} 
The trend of the emission wavelength evolution for different values of the applied voltages when taking all voltage combinations in a $0.5$\,V region. The black lines indicate the slope which the wavelength position varies. The slope of the wavelength position variations is different for different values of the applied voltage. }
   \end{figure}

To determine the underlying cause of the extra phase between the AMZI arms we investigate the response of each EOPM of the AMZI with the applied voltage. For that, we select three different regions where the voltages are switched in a window of $0.5$\,V for different values of $V_\textrm{EOPM1}$ and $V_\textrm{EOPM2}$. We record the optical spectrum for each case with a resolution of $40$\,fm and extract the wavelength position. The data are plotted in Fig.\,\ref{fig:Gradient}, each map represents a different wavelength interval as the applied voltages are different for each case. We identify a trend on the wavelength evolution as the ($V_\textrm{EOPM1},V_\textrm{EOPM2}$) voltage space is scanned, and remark that the slope of the color gradient varies from map to map. When the phase change is the same in both arms of the AMZI the slope should be equal to -$1$, unlike the slope we measure which suggests that the response of the two EOPMs with the applied voltage is different and in addition, also varies with the value of the applied voltage. This nonlinearity results in an extra phase term which slightly shifts the cavity modes, and hence the emission wavelength, as observed in the experiment. In order to successfully employ this tuning approach and correct for the extra phase, a different voltage coefficient should be applied on one EOPM to remove undesired intra-cavity phase variations. Additionally, the voltage correction should be changed approximately every $0.5$\,V.

\section{Control strategy for a semi-continuous wavelength tuning}
\label{sec:tuning strategy}
   \begin{figure}
   \begin{center}
   \includegraphics[width=0.86\linewidth]{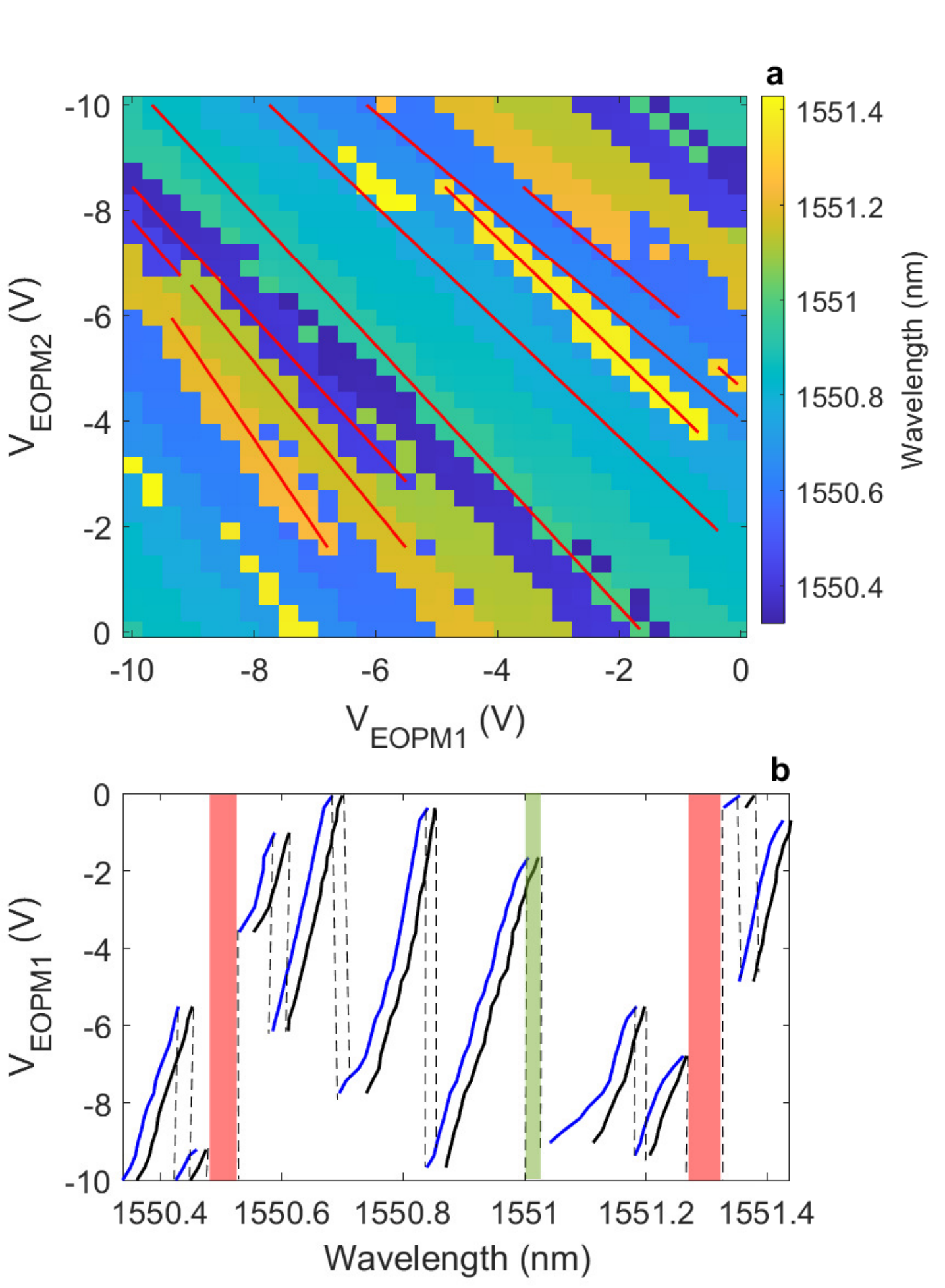}  
   \end{center}
   \caption[Tuning] 
   { \label{fig:Tuning} 
(a) The emission wavelength as a function of the applied voltages in the EOPMs of the AMZI. The voltages are switched from -$10$\,V to $0$\,V in steps of $0.3$\,V while the voltage of the phase section is kept constant at $0$\,V. The red lines represent the segments where the wavelength changes continuously with the applied voltage. (b) The values of the wavelength for each of the selected segments in red in panel (a) as a function of the voltage applied to EOPM1. The black curves show the selected segments when the same measurement is repeated and the voltage applied to the phase section is -$5$\,V. With green is highlighted the region where the phase section can bridge the gap between two consecutive segments and with red the regions that are not accessible with this tuning method.}
   \end{figure} 

To avoid the issues discussed in the previous section, we investigate another approach on controlling the AMZI and tuning the emission wavelength of the laser. Here, instead of correcting for the phase change between the AMZI arms, our goal is to take advantage of the phase difference and shift the selected cavity mode rather then keeping them fixed. For that we identify regions of the parameter space in which the wavelength can be tuned continuously.

To find the regions where the wavelength changes continuously, we build a map with the wavelength positions obtained from all the voltage combinations when the two EOPMs are switched from -$10$\,V to $0$\,V in steps of $0.3$\,V while the voltage of the phase section is kept constant at $0$\,V. The optical spectra are recorded with $1.12$\,pm resolution and the results are plotted in Fig.\,\ref{fig:Tuning}(a). Each color in the map correspond to a different cavity mode and within each color region we notice a gradient, i.e., a wavelength variation, resulting from the phase difference. In the map we locate the sections where the wavelength changes continuously and that reach as many wavelengths as possible. $10$ different segments with length in the order of tens of \,pm are identified, all these segments together can cover the entire FSR range of the AMZI and $89.5$\% of the tuning range of the laser. Each of the selected segments are shown in the map with red lines, their slope changes slightly from one color region to another, as discussed in the previous section. For each of the selected segments we extract the values of the voltages for each EOPM at the start and at the end of the segment. In between these values the voltages are switched continuously which leads to a continuous shift of the selected mode. The position of the emission wavelength variation as a function of the applied voltage is shown in Fig.\,\ref{fig:Tuning}(b) for $V_\textrm{EOPM1}$. A similar plot can be build for the $V_\textrm{EOPM2}$. We notice that there are two regions, highlighted with red in  Fig.\,\ref{fig:Tuning}(b) ($1550.476$\,nm-$1550.530$\,nm and $1551.266$\,nm-$1551.327$\,nm) that are not covered by our tuning method. We can however extend the laser tuning range by employing the phase section. With black in Fig.\,\ref{fig:Tuning}(b) it is shown how the continuous wavelength segments are shifted when -$5$\,V is applied on the phase section. We notice that applying a voltage on the phase section can bridge the gap between two consecutive segments as highlighted with green in Fig.\,\ref{fig:Tuning}(b). For the regions highlighted with red the voltage applied on the phase section narrows the gaps but is not sufficient to completely close them. 

   \begin{figure} 
   \begin{center}
   \includegraphics[width=\linewidth]{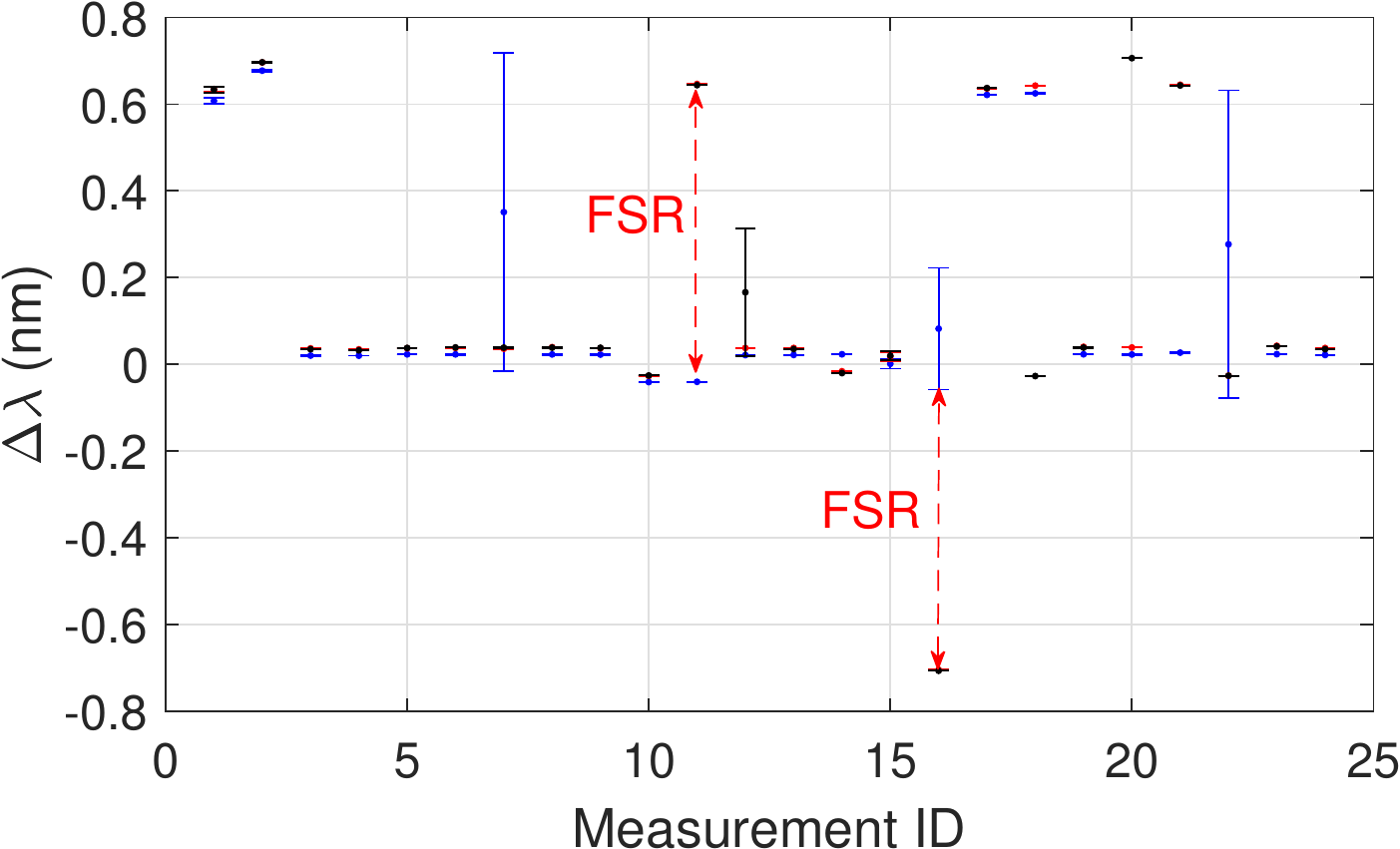}  
   \end{center}
   \caption[Repeatability] 
   { \label{fig:Repeatability} 
The change on the lasing wavelength for $24$ voltage setting configurations measured every $24$\,h for $4$ consecutive days. (Day $2$ - Day $1$) - blue; (Day $3$ - Day $1$) - green; (Day $4$ - Day $1$) - black; The dashed red arrows show the voltage combinations that some days lead to a wavelength change equal to the FSR of the AMZI.}
   \end{figure} 

There are two aspect of this control mechanism that should be considered, the tuning resolution and the measurement repeatability. Since within the selected segments the control voltages are varied continuously, the wavelength tuning resolution will be determined by the step of the voltage variation. For the voltage source that we use (NI9253), the voltage setting resolution is $0.001$\,V. To estimate the wavelength resolution when the voltage changes each step by $0.001$\,V, we measure the variations in the beating frequency between the lasers wavelength and a local oscillator (Keysight N7776C) via a signal analyzer (Keysight N9021B). We found that the wavelength resolution for this control mechanism is $0.1$\,pm when the control voltages are switched by $0.001$\,V. To examine the repeatability of the measurements we select a set of $24$ voltage combinations in the different continuous segments and measure the resulting emission wavelength every $24$\,h across different days. The differences of the emission wavelength measured each day with the wavelength measured on the first day are plotted in Fig.\,\ref{fig:Repeatability} for four consecutive days. We notice that the wavelength varies slightly from day to day and the change in the wavelength is in the range of $\approx\pm 0.02$\,nm. In addition, in some days the wavelength jumps by the FSR of the AMZI, as shown with the dashed red arrows in Fig.\,\ref{fig:Repeatability}. Thus, if wavelength settings more precise that $0.02$\,nm are needed, the calibration step should be repeated before each measurement. 

Altogether, using this control strategy the FSR wavelength jumps are avoided, and the resolution of the wavelength tuning is determined by the voltage setting and not by the filtering mechanism and is able to reach up until sub-\,pm range. The gaps are still present however, they can also be narrowed when the phase section is used. In contrast with the previous method where the value of the $V_\textrm{Phase}$ should be precisely defined in each step, here, the $V_\textrm{Phase}$ is linearly increased until the following continuous region is reached, as illustrated with green in Fig.\,\ref{fig:Tuning}(b). All that is required are the start and end values of the voltages for each of the continuous segments which can be identified by a relatively simple calibration step. After the values of the voltages for each segment are correctly set from a look-up table, the wavelength can be tuned continuously over the biggest part of the tuning range of the laser, on what we refer to as a semi-continuous wavelength tuning.

\section{Conclusion}
We have presented experimental results of the dependence of the emission wavelength of a tunable laser on the tuning of an intra-cavity asymmetric Mach-Zehnder interferometer. The proposed design allows for the thorough study of the filtering properties of an AMZI, hence, we can tackle the challenges related to the robust control of its filtering mechanism. The laser considered in this work features a tuning range of about $1.2$\,nm with wavelength switching speed between $60$\,ns and $80$\,ns, an intrinsic linewidth of $381$\,kHz and an SMSR over $20$\,dB over the whole range.

We have investigated the control efficiency of the EOPMs comprising the AMZI and we found that the performance of the EOPMs on the same chip varies and the phase shift of the EOPMs exhibits a nonlinear behavior with the applied voltage resulting in an extra phase between the AMZI arms. Hence, selecting consecutive longitudinal modes by simply linearly changing the voltages applied to the EOPMs becomes rather challenging. Instead, we propose a control strategy that translates continuously the cavity modes in a certain range rather then selecting consecutively the cavity modes. For this strategy the voltage combinations that select specific cavity modes within the tuning range are necessary and could be found by a simple calibration step in order to obtain coverage of the entire FSR of the AMZI and the biggest part of the tuning range, thus, we refer to this strategy as a semi-continuous wavelength tuning. The wavelength resolution on the proposed control strategy depends on the accuracy of the voltage setting and is usually in the order of sub-pm. The limited tuning range constrains the use of this laser in wide range of application although it could be advantageously used on applications that require high resolution and precise wavelength control. 


\ifCLASSOPTIONcaptionsoff
  \newpage
\fi

\bibliography{IEEEabrv, report}
\bibliographystyle{IEEEtran}

\end{document}